\documentclass[11pt,a4paper]{article}
\usepackage{amsmath}
\usepackage{amssymb}
\usepackage{graphicx}
\newcommand {\be} {\begin{eqnarray*}}
\newcommand {\ee} {\end{eqnarray*}}
\newcommand {\bea} {\begin{eqnarray}}
\newcommand {\eea} {\end{eqnarray}}
\newcommand \foot[1]{\footnotemark\footnotetext{#1}}
\begin{document}
\begin{center}
{\Large{\textbf{The Big Bang as a Phase Transition}}}
\end{center}

\begin{center}
\textbf{Tom$\acute{\mbox{a}}\check{\mbox{s}}$ Liko and Paul S. Wesson} \\
{\small Department of Physics, University of Waterloo} \\
{\small Waterloo, Ontario, Canada, N2L 3G1} \\
{\small tliko@uwaterloo.ca}
\end{center}

\begin{abstract}
We study a five-dimensional cosmological model, which suggests that the universe
began as a discontinuity in a scalar (Higgs-type) field, or alternatively as a
conventional four-dimensional phase transition.
\end{abstract}

\section{Introduction}
Recent years have witnessed a large amount of interest in higher-dimensional cosmologies
where the extra dimensions are noncompact.  A popular example is the so-called
Randall-Sundrum braneworld scenerio \cite{rs1, rs2}.  This is typically a five-dimensional
$\mbox{AdS}_{5}$ black hole spacetime called the bulk, on which our universe is described
by a domain wall called the brane.  All the matter interactions are assumed to be confined
to this brane, except for gravitation, which is allowed to propagate in the bulk.  This
model was inspired by a certain string theory discovered by Ho$\check{\mbox{r}}$ava and
Witten \cite{hw1, hw2} who showed that extra dimensions did not need to be compactified.
In this scenario, the fields of the standard model are represented by strings whose endpoints
reside on a 10D hypersurface, and are therefore confined to this ``brane''.  The
gravitational degrees of freedom, on the other hand, are represented by closed strings
which is why the graviton is free to propagate along the entire 11D manifold.  Prime among
all these models is the concept of $\mathbb{Z}_{2}$ symmetry, which is simply the requirement
that the manifold on one side of the brane be the mirror image of the other side.  An
alternative approach which uses a large extra dimension is Space-Time-Matter, proposed by
Wesson, which is so-called because the foundation is that the fifth dimension induces
matter \cite{wesson1}.  This point of view
introduces a certain symmetry in physics, since in mechanics we normally use as our base
dimensions length, time, and mass.  A realization of this theory is that all the matter
fields in 4D can arise from a higher-dimensional vacuum.  One starts with the vacuum Einstein
field equations in 5D, and dimensional reduction of the Ricci tensor leads to an effective
4D energy-momentum tensor \cite{wesleon}.  For this reason the theory is also called
induced-matter theory.  These two theories (braneworld and induced-matter) may apprear to be
different, but have recently been shown to be equivalent by Ponce de Leon
\cite{leon}.  The aim of this paper is to demonstrate how a solution of induced-matter
theory can be used to generate a simple braneworld cosmology with $\mathbb{Z}_{2}$ symmetry,
and to study the properties of this cosmology, paying particular attention to a phase
transition which occurs at the big bang.

We begin by writing down a class of exact cosmological solutions to the 5D vacuum
(Ricci-flat) field equations $R_{AB}=0$ \cite{liuwes, xlw, wlx}.  The 5D line
element is:
\bea
dS^{2} &=& \frac{\dot{a}^{2}}{\mu^{2}}dt^{2}
           - a^{2}\left(\frac{dr^{2}}{1-kr^{2}}+r^{2}d\Omega^{2}\right)
           - dy^{2}\nonumber\\
a^{2} &=& (\mu^{2}+k)y^{2} + 2\nu y + \frac{\nu^{2}+K}{\mu^{2}+k}.
\label{lmw1}
\eea
Here the functions $\mu\equiv\mu(t)$ and $\nu\equiv\nu(t)$, $\dot{a}=\partial a/\partial t$ and
$k$ $(+1,0,-1)$ and $K$ are constants.\foot{\emph{Conventions:}  In this paper we use upper-case
latin indices to run $0,\ldots,4$, lower-case greek indices to run $0,\ldots,3$, and the
signature of the metric is always $(+-...-)$.  We use spherical coordinates
$x^{01234}=tr\theta\phi y$ with $d\Omega^{2}=d\theta^{2}+\sin^{2}\theta d\phi^{2}$.  Unless
otherwise stated, we work in natural units where $c=8\pi G=1$.}  Studies of this class of
solutions show that it is algebraically broad, and depending on the choice of coordinates
has two feasible interpretations: (a) a hot early universe, with matter production typical
of inflationary quantum field theories, and a decaying cosmological constant of the kind
needed to resolve the timescale problem of standard cosmology \cite{liuwes}; (b) a 5D
topological black hole (see below) \cite{seahrathesis, seawes1}.  The constant $K$ appearing
in (\ref{lmw1}) is a constant of integration and is related to the 5D Kretchmann scalar via
\bea
R_{ABCD}R^{ABCD} = \frac{72K^{2}}{a^{8}},
\label{k5}
\eea
which is the only geometrical invariant that is non-zero since $R_{AB}R^{AB}=0$ and
$R=0$ by the field equations.  It should be noted that $a=0$ corresponds to a
geometrical singularity for the 5D model which is similar to those that occur in
the 4D Friedmann-Robertson-Walker (FRW) models.  In fact, it can be shown via a
non-trivial coordinate transformation $R=R(t,y)$, $T=T(t,y)$ that (\ref{lmw1}) is
isometric to the (topological) black hole manifold with line element
\bea
d\mathcal{S}^{2} = h(R)dT^{2} - h^{-1}(R)dR^{2} - R^{2}\left[d\psi^{2}
                   + S_{k}^{2}(\psi)(d\theta^{2} + \sin^{2}\theta d\phi^{2})\right], 
\eea
where $h(R)=k-K/R^{2}$, and the function
\bea
S_{k}(\psi) = 
\left\{
\begin{array}{ll}
\sin\psi,\phantom{a} & k = +1,\\
\psi,\phantom{aaaa} & k = 0,\phantom{+}\\
\sinh\psi, & k = -1.
\end{array}
\right.
\eea
In what follows, we are not concerned with this singularity.  Instead we consider a
different type of singularity that occurs at $\dot{a}/\mu=0$, corresponding to a
coordinate singularity similar to that which defines an event horizon, but now the
principal pressures of the material fluid also diverge at this point.  This part of
the manifold thus defines a matter singularity.

\section{A Simple Cosmological Model}
We now proceed to illustrate how the solution (\ref{lmw1}) can be used to generate a
braneworld model.  In what follows, we appeal to astrophysical data so we set $k=0$,
and for simplicity choose $K=1$.  We are also interested in a solution which incorporates
the $\mathbb{Z}_{2}$ reflection symmetry condition
$g_{\alpha\beta}(x^{\gamma},y)=g_{\alpha\beta}(x^{\gamma},-y)$.  This is achieved
simply by setting $\nu=0$, so the line element (\ref{lmw1}) takes the form
\bea
dS^{2} &=& \frac{\dot{a}^{2}}{\mu^{2}}dt^{2} - a^{2}(dr^{2}+r^{2}d\Omega^{2})
           - dy^{2}\nonumber\\
a^{2} &=& \mu^{2}y^{2} + \frac{1}{\mu^{2}}\nonumber\\
\frac{\dot{a}^{2}}{\mu^{2}} &=& \frac{(y^{2}-\mu^{-4})^{2}}{(y^{2}+\mu^{-4})}
                                \frac{1}{\mu^{2}}\left(\frac{d\mu}{dt}\right)^{2}.
\label{lmw2}
\eea
In order to make contact with the matter properties that this solution describes,
we make note of Campbell's embedding theorem which states that any analytic
$(N-1)$D Riemannian manifold can be locally embedded in an $N$D Riemannian manifold
that is Ricci-flat \cite{campbell, rrt, rtz, lrtr, seawes2}.  This provides a basis
for interpreting a solution of $R_{AB}=0$ like (\ref{lmw2}) as a solution of the
4D field equations $G_{\alpha\beta}=T_{\alpha\beta}$ with sources.

The functional form of the stress-energy tensor
$T_{\alpha\beta}=T_{\alpha\beta}(x^{\gamma},y)$ has been known for some years
\cite{wesleon} and is:
\bea
T_{\alpha\beta}
&=& \frac{\nabla_{\beta}(\partial_{\alpha}\Phi)}{\Phi}
- \frac{\varepsilon}{2\Phi^{2}}\Biggl\lbrace\frac{\stackrel{*}{\Phi}
    \stackrel{*}{g}_{\alpha\beta}}{\Phi}
- \stackrel{**}{g}_{\alpha\beta}
+ g^{\lambda\mu}\stackrel{*}{g}_{\alpha\lambda}\stackrel{*}{g}_{\beta\mu}\nonumber\\
&\phantom{=}& - \frac{1}{2}g^{\mu\nu}\stackrel{*}{g}_{\mu\nu}\stackrel{*}{g}_{\alpha\beta}
+ \frac{1}{4}g_{\alpha\beta}\left[\stackrel{*}{g}^{\mu\nu}\stackrel{*}{g}_{\mu\nu}
+ \left(g^{\mu\nu}\stackrel{*}{g}_{\mu\nu}\right)^{2}\right]\Biggr\rbrace.
\label{effect}
\eea
Here, $\nabla_{\beta}$ denotes the covariant derivative,
$\partial_{\beta}\equiv\partial/\partial x^{\beta}$, and the overstar denotes partial
differentiation with respect to the fifth coordinate.  Also, the fifth component of the
metric is $\epsilon\Phi^{2}$, where $\epsilon=\pm1$ and $\Phi$ is a scalar field which may
be related to particle mass \cite{wesson2, wesson3}.  Campbell's theorem can in fact be
inferred from the ADM formalism, which has been used to obtain the 4D energy of 5D solutions
\cite{bkkls, sw}.  The embedding expressions need modification if
there is a singular surface, as may happen in $\mathbb{Z}_{2}$-symmetric cosmologies; but there
is mathematical consistency between induced-matter and brane models \cite{leon}.  Here
we would like to emphasize that while complementary techniques may be employed to obtain
a unique functional form for the stress-energy tensor, there is still an ambiguity
involved in the physical interpretation of this.  An analogous situation occurs in
4D, where a given metric may have different sources.  As in the standard FRW models,
we assume here that the source is a perfect fluid with energy density $\rho$ and pressure
$p$ so that
\bea
T_{\alpha\beta} = (\rho + p)u_{\alpha}u_{\beta} - pg_{\alpha\beta}.
\label{setensor}
\eea
This stress-energy tensor must satisfy the 4D field equations with a line element
$ds^{2}=g_{\alpha\beta}dx^{\alpha}dx^{\beta}$ which is the 4D part of (\ref{lmw2}).
If we take the matter to be comoving, then the four-velocities
$u^{\alpha}=dx^{\alpha}/ds$ satisfy $u^{\alpha}=(u^{0},0,0,0)$ and $u^{0}u_{0}=1$.
Then we find
\bea
\rho
&=& \frac{3\mu^{4}}{\mu^{4}y^{2}+1}\nonumber\\
p
&=& -\mu^{4}\left(\frac{2}{\mu^{4}y^{2}-1}+\frac{1}{\mu^{4}y^{2}+1}\right).
\label{stress1}
\eea
Note that the pressure is generally discontinuous at the point $\mu^{2} = 1/y$.
This is also the point at which the scale factor $a$ is a minimum.

An examination of (\ref{lmw1}) and (\ref{lmw2}) shows that the form of
$(\dot{a}/\mu)dt$ is invariant under an arbitrary coordinate transformation
$t\rightarrow t(\tilde{t})$.  As a result, we can freely choose the form of
$\mu(t)$ without loss of generality, and it will prove convenient to take
$\mu(t)=1/\sqrt{2t}$.  From (\ref{lmw2}) we find that
\bea
dS^{2} = \left(1-\frac{y^{2}}{4t^{2}}\right)^{2}\left(1+\frac{y^{2}}{4t^{2}}\right)^{-1}
         dt^{2} - \left(2t+\frac{y^{2}}{2t}\right)(dr^{2}+r^{2}d\Omega^{2})
           - dy^{2}.
\label{lmw3}
\eea
This shows that the scale factor for the 3D sections of the 5D metric has the
time-dependence typical of the radiation-dominated FRW model, but with an extra term
that comes from dependence on the fifth coordinate.  This term decays with coordinate
time $t$.  Alternatively, the fifth dimension should be important as $t\rightarrow0$.
Since $\dot{a}/\mu\rightarrow1$ as $t\rightarrow\infty$, the coordinate time becomes the
cosmic time.  Note that this solution describes the 4D radiation FRW model on the
surface of reflection ($y=0$).  For general $y\neq0$ and $t\rightarrow\infty$
we have $a\rightarrow\sqrt{2t}$ and $\dot{a}/\mu\rightarrow1$.  Then
\bea
dS^{2} \rightarrow dt^{2} - 2t(dr^{2}+r^{2}d\Omega^{2}) -dy^{2}
\quad
\mbox{as}
\quad
t \rightarrow \infty,
\eea
which is the embedded radiation model.  For general $y\neq0$ and $t\rightarrow0$
we have $a\rightarrow y/\sqrt{2t}$ and $\dot{a}/\mu\rightarrow y/2t$.  This gives
\bea
dS^{2} \rightarrow \frac{y^{2}}{L^{2}}\left[\left(\frac{L}{2t}\right)^{2}dt^{2}
                   -\frac{L^{2}}{2t}(dr^{2}+r^{2}d\Omega^{2})\right] - dy^{2}
\quad
\mbox{as}
\quad
t \rightarrow 0,
\label{canonical}
\eea
where we have introduced a constant length $L$ to make contact with 5D metrics in the
canonical form \cite{wesson4}.  This has the general form
\bea
dS^{2} = \frac{y^{2}}{L^{2}}[\bar{g}_{\alpha\beta}(x^{\gamma},y)dx^{\alpha}dx^{\beta}]
         - dy^{2},
\eea
and is useful because its first part can be related to the 4D action of quantum
physics and leads to great simplification of the geodesic equation of classical
physics.  Both brane models and induced-matter theory lead in general to a fifth
force which is zero if $\partial\bar{g}_{\alpha\beta}/\partial y=0$.  This can be
physically justified by appeal to the weak equivalence principle (which is then a
symmetry of the metric).  Alternatively, it is mathematically justified by the
argument that the metric not be significantly affected by the mass of a test particle
which moves through it (no reaction force).  The noted condition is satisfied by
(\ref{canonical}).  This via $t=e^{2\tau/L}$ tranforms to
\bea
dS^{2} \rightarrow \frac{y^{2}}{L^{2}}\left[d\tau^{2}
                   - \frac{1}{2}L^{2}e^{-2\tau/L}(dr^{2}+r^{2}d\Omega^{2})\right]
                   - dy^{2}
\quad
\mbox{as}
\quad
\tau \rightarrow -\infty,
\eea
whose 4D part is a de Sitter space of the kind used in other applications of brane theory.

Turning our attention to the induced matter described by (\ref{lmw3}), we find from
(\ref{stress1}) that
\bea
\rho
&=& \frac{3}{4t^{2}+y^{2}}\nonumber\\
p
&=& \frac{2}{4t^{2}-y^{2}}-\frac{1}{4t^{2}+y^{2}}.
\label{stress2}
\eea
From these comes the inertial density of matter:
\bea
\rho+p = \frac{16t^{2}}{(4t^{2}+y^{2})(4t^{2}-y^{2})},
\label{idensity}
\eea
which is regarded as setting a condition for the stability of matter via
$\rho+p>0$.  From (\ref{stress2}) also comes the gravitational density of matter:
\bea
\rho+3p = \frac{6}{4t^{2}-y^{2}},
\label{gdensity}
\eea
which is regarded as setting a condition for the gravity of matter via
$\rho+3p>0$.  For the case of (\ref{lmw3}) this means that
$t>y/2$.  The form of relations (\ref{stress2}) shows that the energy
density and pressure vary differently with $t$ and $y$.  The behaviour of
these is summarized in Table 1.
\begin{table}[t] 
\begin{tabular*}{\linewidth}{@{\extracolsep{\fill}}lcccc} 
\hline
\hline
\phantom{t} & \quad $\rho$ & \quad $p$ & \quad $\rho+p$ & \quad $\rho+3p$ \\
\hline
& & \\
$t\rightarrow0$ & \quad $\frac{3}{y^{2}}$ & \quad $-\frac{3}{y^{2}}$ &
\quad $0$ & \quad $-\frac{6}{y^{2}}$ \\
& & \\
$t\rightarrow\frac{y}{2}^{-}$ & \quad $\frac{3}{2y^{2}}$ &
       \quad $-\infty$ & \quad $-\infty$ & \quad $-\infty$ \\
& & \\
$t\rightarrow\frac{y}{2}^{+}$ & \quad $\frac{3}{2y^{2}}$ & 
         \quad $+\infty$ & \quad $+\infty$ & \quad $+\infty$ \\
& & \\
$t\rightarrow\infty$ & \quad $\frac{3}{4t^{2}}$ & \quad $\frac{1}{4t^{2}}$ &
\quad $\frac{1}{t^{2}}$ & \quad $\frac{3}{2t^{2}}$ \\
& & \\
\hline
\hline
\end{tabular*}
\caption{Asymptotic behaviour of the stress-energy, from (\ref{stress2})-(\ref{gdensity}).}
\end{table}

The universe described by (\ref{lmw3}) begins in an infinitely distended state
with the equation of state $\rho+p=0$, typical of 5D vacuum cosmologies.  It contracts,
with the energy density decaying from its maximum by $50\%$ and the pressure becoming
unbounded as $t\rightarrow y/2$.  The 5D Kretchmann scalar (\ref{k5}) remains finite
and in fact drops to its minimum $9/2y^{4}$ at this point.  The pressure changes
discontinuously as in a phase transition.  After this, the universe expands with the
matter becoming radiation-like, and the energy density decreasing in the same way as in
standard ($k=0$) cosmology.  This behaviour of the induced matter is shown in Fig. 1,
where we have set $y=1$.
\begin{figure}[t]
\centering
\includegraphics[width=2.1in]{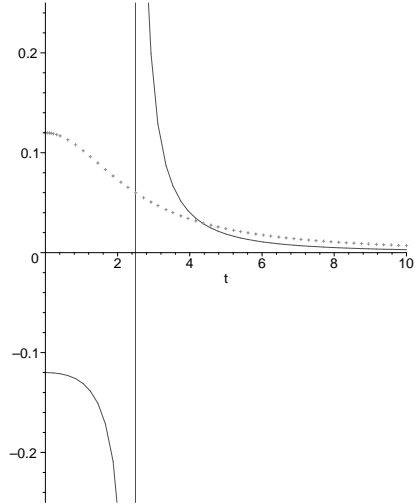}
\caption{Evolution of the energy density and pressure with time, from (\ref{stress2}).
The dotted line represents the energy density and the solid line represents the pressure.}
\end{figure}

It should be noted that we have considered the stress-energy for general $y\neq0$
hypersurfaces, but the surface of reflection $y=0$ defines a special hypersurface
on the manifold.  For here there is no phase transition, and we have $a\rightarrow0$
and $\rho\rightarrow\infty$ for $t\rightarrow0$, thus defining a big bang.  By
(\ref{stress2}) we have $\rho = 3/4t^{2}$ and $p = 1/4t^{2}$, which implies a
radiation-dominated ($k=0$) cosmology.  We may also consider the energy density
$\rho(t,y)=3/(4t^{2}+y^{2})$ in (\ref{stress2}).  There is a maximum at the
$y=0$ hypersurface, and $\rho$ is not uniformly distributed along the
$y$-direction.  In fact, this kind of concentration can be seen more clearly in the
early stage of the universe where $t\sim y$, as is shown in Fig. 2.
\begin{figure}[t]
\centering
\includegraphics[width=2.5in]{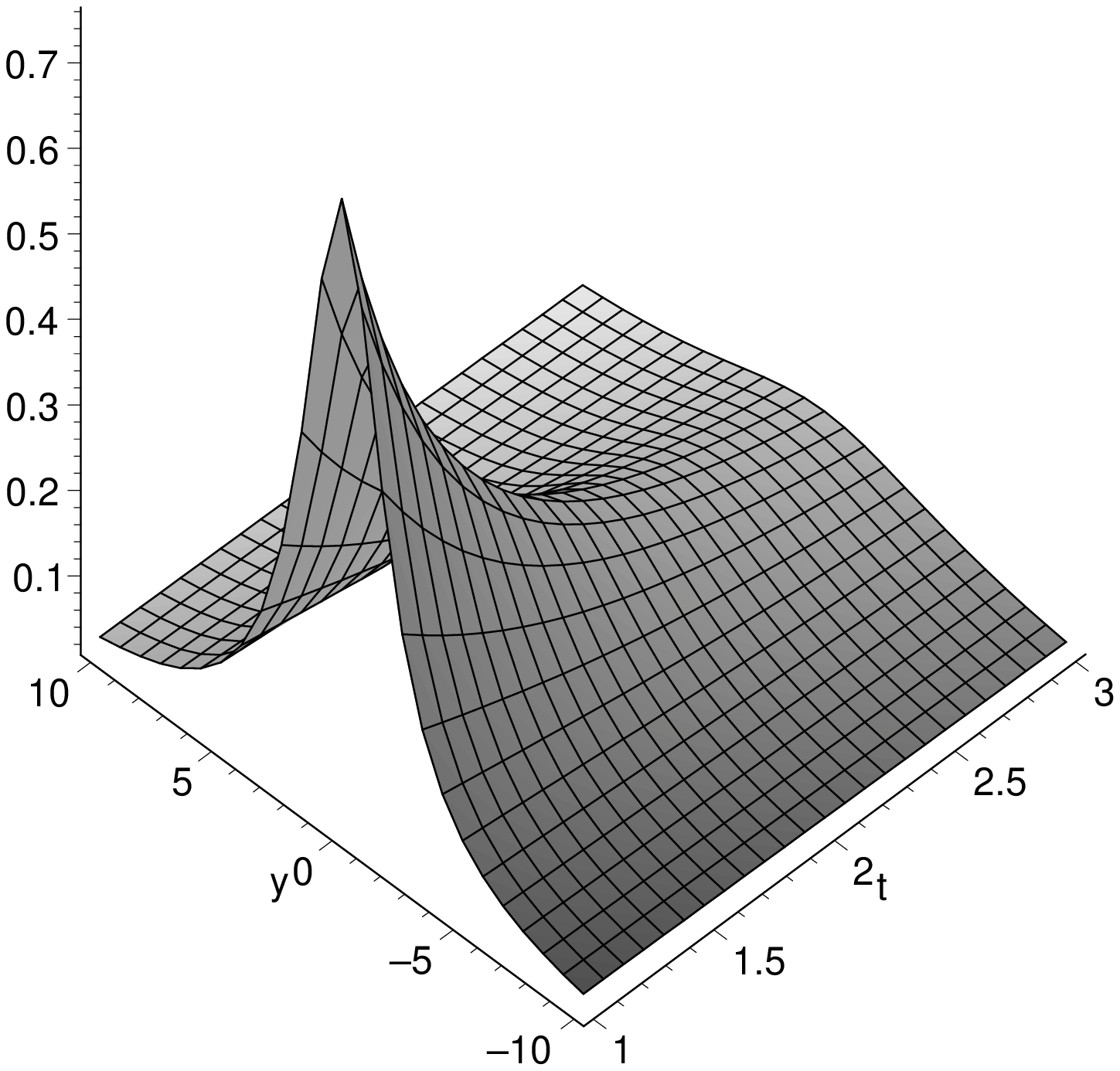}
\caption{Distribution of the energy density in the early universe, from (\ref{stress2}).}
\end{figure}

\section{Discussion}
We have investigated in detail the stress-energy behaviour of the matter that is
associated with the solution (\ref{lmw3}), but the source of the phase transition is
currently unknown.   Also, as mentioned above, the nature of this matter is open to
interpretation.  For example, we can decompose (\ref{stress2}) into a linear combination
of two stress-energy tensors such that
\bea
T_{\beta}^{\alpha} &=& \left(T_{\beta}^{\alpha}\right)_{I}
                + \left(T_{\beta}^{\alpha}\right)_{II}\nonumber\\
\left(T_{\beta}^{\alpha}\right)_{I} &=& \left[\frac{1}{4t^{2}+y^{2}}\right]
                                 \mbox{diag}(3,1,1,1)\nonumber\\
\left(T_{\beta}^{\alpha}\right)_{II} &=& -\left[\frac{2}{4t^{2}-y^{2}}\right]
                                 \mbox{diag}(0,1,1,1).
\label{decomposition}
\eea
The first part $(T_{\beta}^{\alpha})_{I}$ has the equation of state $\rho+3p=0$
which describes non-gravitating matter.  The second part $(T_{\beta}^{\alpha})_{II}$
has the functional form $f(t)\mbox{diag}(0,1,1,1)$ on a $y=\mbox{constant}$
hypersurface.  It is this second component which is the source associated with the
phase transition at $t=y/2$.  The decomposition (\ref{decomposition}) has two
possible interpretations: (a) a universe that is filled with two different types of
(possibly non-interacting) matter fields; (b) a universe filled with a non-gravitating
matter field subject to a bulk viscosity. 

In either physical case, there is a corresponding geometrical interpretation.
Recall that the 5D Kretchmann scalar remains finite across the boundary $t=y/2$.
On the other hand, its 4D counterpart given by
\bea
R_{\alpha\beta\gamma\delta}R^{\alpha\beta\gamma\delta}
= \frac{24}{(4t^{2}+y^{2})(4t^{2}-y^{2})}
\eea
becomes singular.  This would suggest that the discontinuity in the pressure is really
a manifestation of a cusp in the 4D geometry.  However, the relations (\ref{idensity})
and (\ref{gdensity}) for the inertial and gravitational densities of the matter need to
be taken into consideration.  The stability of these requires that $t>y/2$, but the
phase transition occurs \emph{at} $t=y/2$.  Thus, the solution (\ref{lmw3}) as
illustrated in Fig. 1 has a period when the matter is exotic; and after the transition
at $t=y/2$ a period where it is normal and indeed asymptotic to a photon gas.

Let us take a closer look at the ``matter'' of the model.  We have already commented that
there is a discontinuity in the pressure $p$, which by the laws of 4D thermodynamics would
lead us to identify the event at $t=y/2$ as a first-order phase transition.  However, the
matter described by (\ref{stress2}) or (\ref{decomposition}) before the transition is exotic,
insofar as it violates the usual energy conditions.  This suggests to us that, while our
model is classical, the underlying mechanism for the transition has to do with quantum
effects.  To illustrate this, we note that we can model the matter in (\ref{stress2}) or
(\ref{decomposition}) by a scalar field $\Psi=\Psi(x^{A})$.  The correspondence
between the classical and quantum approaches is established by matching the classical
stress-energy tensor (\ref{setensor}) to the equivalent relation for the quantum-field
theory expression for a scalar field:
\bea
T_{\alpha\beta} = \partial_{\alpha}\Psi\partial_{\beta}\Psi - g_{\alpha\beta}\mathcal{L}.
\eea
Here $\mathcal{L}$ is the Lagrangian density
\bea
\mathcal{L} = \frac{1}{2}\partial_{\alpha}\Psi\partial^{\alpha}\Psi - V(\Psi),
\eea
and $V(\Psi)$ is an arbitrary potential.  Then we find that the classical expressions for
the density and pressure are given by
\bea
\rho &=& \frac{1}{2}\dot{\Psi}^{2} + \frac{1}{2}(\nabla\Psi)^{2} + V(\Psi)\nonumber\\
p &=& \frac{1}{2}\dot{\Psi}^{2} - \frac{1}{6}(\nabla\Psi)^{2} - V(\Psi).
\label{odes}
\eea
These expressions are generic.  However, modulo the non-uniqueness of the source responsible
for (\ref{lmw3}), the phase transition in (\ref{stress2}) corresponds to a discontinuity
in a Higgs-type field.

\section{Conclusion}
There are many models for the ``big bang'', but they suffer from the problem that the
origin of the matter is unexplained.  We have given in the above a simple model.  In a
classical sense, the big bang is a first-order thermodynamic phase transition.  In a
quantum sense, it corresponds to a discontinuity in a Higgs-type scalar field.  This
picture is conformable with other models, including those of Vilenkin \cite{vilenkin},
where the ``big-bang'' is a tunelling event.  We cannot completely identify the nature
of this event, given the latitude inherent in the Lagrangian which describes a quantum
scalar field.  However, we feel that the description in terms of classical fields is an
improvement over the traditional one: if the big bang is a phase transition, it opens the
way to further thermodynamic investigations.

\section*{Acknowlegdements}
We thank S.S. Seahra for discussions, and R.C. Myers for commenting on an earlier draft
of the manuscript.  This work grew out of an earlier collaboration with H. Liu.  This
work was supported in part by N.S.E.R.C. grant number 200917.


\begin{thebibliography}{99}
\bibitem{rs1}
Randall, L., Sundrum, R. 1999, ``A Large Mass Hierarchy from a Small Extra Dimension.''
\emph{Phys. Rev. Lett.} \textbf{83}, 3370. hep-th/9905221.
\bibitem{rs2}
Randall, L., Sundrum, R. 1999, ``An Alternative to Compactification.''
\emph{Phys. Rev. Lett.} \textbf{83}, 4690. hep-th/9906064.
\bibitem{hw1}
Ho$\check{\mbox{r}}$ava, P., Witten, E. 1996, ``Heterotic and Type I String Dynamics
from Eleven Dimensions.''
\emph{Nucl. Phys. B.} \textbf{460}, 506. hep-th/9510209.
\bibitem{hw2}
Ho$\check{\mbox{r}}$ava, P., Witten, E. 1996, ``Eleven-Dimensional Supergravity
on a Manifold with Boundary.''
\emph{Nucl. Phys. B.} \textbf{475}, 94. hep-th/9603142.
\bibitem{wesson1}
Wesson, P.S. 1984, ``An Embedding for General Relativity with Variable Rest Mass.''
\emph{Gen. Rel. Grav.} \textbf{16}, 193.
\bibitem{wesleon}
Wesson, P.S., Ponce de Leon, J. 1992, ``Kaluza-Klein Equations, Einstein's Equations,
and an Effective Energy-Momentum Tensor.''
\emph{J. Math. Phys.} \textbf{33}, 3883.
\bibitem{leon}
Ponce de Leon, J. 2001, ``Equivalence Between Space-Time-Matter and Brane-World Theories.''
\emph{Mod. Phys. Lett. A.} \textbf{16}, 2291. gr-qc/0111011.
\bibitem{liuwes}
Liu, H., Wesson, P.S. 2001, ``Universe Models with a Variable Cosmological `Constant' and
a `Big Bounce'.''
\emph{Astrophys. J.} \textbf{562}, 1. gr-qc/0107093.
\bibitem{xlw}
Xu, L., Liu, H., Wang, B. 2003, ``On the Big Bounce Singularity of a Simple 5D
Cosmological Model.'' \emph{Chin. Phys. Lett.} \textbf{20}, 995. gr-qc/0304049.
\bibitem{wlx}
Wang, B., Liu, H., Xu, L. 2004, ``Accelerating Universe in a Big Bounce Model.''
\emph{Mod. Phys. Lett. A.} \textbf{19}, 449. gr-qc/0304093.
\bibitem{seahrathesis}
Seahra, S.S. 2003, ``Physics in Higher-Dimensional Manifolds.''
PhD Thesis, University of Waterloo.
\bibitem{seawes1}
Seahra, S.S., Wesson, P.S. 2003, ``Universes Encircling 5-Dimensional Black Holes.''
\emph{J. Math. Phys.} \textbf{44}, 5664.  gr-qc/0309006.
\bibitem{campbell}
Campbell, J.E. 1926, \emph{A Course of Differential Geometry}, Clarendon, Oxford.
\bibitem{rrt}
Rippl, S., Romero, C., Tavakol, R. 1995, ``D-Dimensional Gravity from
$(\mbox{D}+1)$-Dimensions.''
\emph{Class. Quant. Grav.} \textbf{12}, 2411. gr-qc/9511016.
\bibitem{rtz}
Romero, C., Tavakol, R., Zalatetdinov, R. 1996, ``The Embedding of General Relativity in
Five-Dimensions.''
\emph{Gen. Rel. Grav}. \textbf{28}, 365.
\bibitem{lrtr}
Lidsey, J.E., Romero, C., Tavakol, R., Rippl, S. 1997, ``On Applications of Campbell's
Embedding Theorem.''
\emph{Class. Quant. Grav.} \textbf{14}, 865. gr-qc/9907040.
\bibitem{seawes2}
Seahra, S.S., Wesson, P.S. 2003, ``Application of the Campbell-Magaard Theorem to
Higher-Dimensional Physics.''
\emph{Class. Quant. Grav.} \textbf{20}, 1321. gr-qc/0302015.
\bibitem{wesson2}
Wesson, P.S. 1999, \emph{Space-Time-Matter}, World Scientific, Singapore.
\bibitem{wesson3}
Wesson, P.S. 2003, ``The Equivalence Principle as a Symmetry.''
\emph{Gen. Rel. Grav.} \textbf{35}, 307. gr-qc/0302092.
\bibitem{bkkls}
Bombelli, L., Koul, R.K., Kunstatter, G., Lee, J., Sorkin, R.D. 1987,
``On Energy in Five-Dimensional Gravity and the Mass of the Kaluza-Klein Monopole.''
\emph{Nucl. Phys. B.} \textbf{289}, 735.
\bibitem{sw}
Sajko, W.N., Wesson, P.S. 2001, ``The Energy of 5D Solitons.''
\emph{Mod. Phys. Lett. A.} \textbf{16}, 627.
\bibitem{wesson4}
Wesson, P.S. 2002, ``Classical and Quantized Aspects of Dynamics in Five-Dimensional
Relativity.''
\emph{Gen. Rel. Grav.} \textbf{19}, 2825. gr-qc/0204048.
\bibitem{vilenkin}
Vilenkin, A. 1982, ``Creation of Universes from Nothing.'' \emph{Phys. Lett. B.}
\textbf{117}, 25.
\end{thebibliography}
\end{document}